\documentclass[sigconf]{acmart}
\usepackage{graphicx}
\usepackage{subcaption}
\usepackage{hyperref}
\usepackage{url}
\usepackage{algorithm}
\usepackage{algorithmic}

\usepackage{amsmath, amsfonts}
\usepackage{bbold}

\newcommand{\EE}{\mathbb{E}}
\newcommand{\RR}{\mathbb{R}}

\newcommand{\closedb}[1]{\left[ #1\right]}
\newcommand{\bs}[1]{\left\{ #1\right\}}

\newcommand{\abs}[1]{\left| #1 \right|}

\usepackage[textsize=tiny,textwidth=1.4cm]{todonotes}

\AtBeginDocument{%
  }

\copyrightyear{2025} 
\acmYear{2025} 
\setcopyright{cc}
\setcctype{by}
\acmConference[ICAIF '25]{6th ACM International Conference on AI in
Finance}{November 15--18, 2025}{Singapore, Singapore}
\acmBooktitle{6th ACM International Conference on AI in Finance (ICAIF '25), November 15--18, 2025, Singapore, Singapore}
\acmDOI{10.1145/3768292.3770344}
\acmISBN{979-8-4007-2220-2/2025/11}

\begin{document}

\title{Extracting the Structure of Press Releases for \\ Predicting Earnings Announcement Returns}

\author{Yuntao Wu}
\email{winstonyt.wu@mail.utoronto.ca}
\orcid{0009-0006-6269-1603}
\affiliation{%
  \institution{University of Toronto}
  \city{Toronto}
  \state{Ontario}
  \country{Canada}
}
\author{Ege Mert Akin}
\email{mert.akin@mail.utoronto.ca}
\orcid{0009-0001-1346-0095}
\affiliation{%
  \institution{University of Toronto}
  \city{Toronto}
  \state{Ontario}
  \country{Canada}
}
\author{Charles Martineau}
\email{charles.martineau@rotman.utoronto.ca}
\orcid{0000-0002-6896-184X}
\affiliation{%
  \institution{University of Toronto}
  \city{Toronto}
  \state{Ontario}
  \country{Canada}
}
\author{Vincent Grégoire}
\email{vincent.gregoire@hec.ca}
\orcid{0000-0002-1230-5026}
\affiliation{%
  \institution{HEC Montréal}
  \city{Montréal}
  \state{Québec}
  \country{Canada}
}
\author{Andreas Veneris}
\email{veneris@eecg.toronto.edu}
\orcid{0000-0002-6309-8821}
\affiliation{%
  \institution{University of Toronto}
  \city{Toronto}
  \state{Ontario}
  \country{Canada}
}

\renewcommand{\shortauthors}{Y. Wu, E. M. Akin, C. Martineau, V. Grégoire, A. Veneris}

\begin{abstract}
We examine how textual features in earnings press releases predict stock returns on earnings announcement days. Using over 138,000 press releases from 2005 to 2023, we compare traditional bag-of-words and BERT-based embeddings. We find that press release content (soft information) is as informative as earnings surprise (hard information), with FinBERT yielding the highest predictive power. Combining models enhances the explanatory strength and interpretability of the content of press releases. Stock prices fully reflect the content of press releases at market open. If press releases are leaked, it offers predictive advantage. Topic analysis reveals self-serving bias in managerial narratives. Our framework supports real-time return prediction through the integration of online learning, provides interpretability and reveals the nuanced role of language in price formation.
\end{abstract}





\maketitle

\section{Introduction}

Earnings announcements are some of the most important anticipated scheduled announcements by corporations in financial markets, often triggering immediate and substantial price adjustments. While research has traditionally focused on the ``hard'' components of earnings releases—namely, the numerical earnings surprises that deviate from analyst expectations—a growing body of research shows that substantial market-relevant information is also embedded in the ``soft'' content of these announcements. In particular, earnings press releases—which are the company-issued quarterly performance statements disseminated as the initial source of public news are a major vehicle through which both hard and soft information reaches investors. These texts often include management’s interpretation of performance, strategic positioning, and forward-looking commentary, all of which can influence investor expectations about the firm's future prospects. 
Despite their importance, earnings press releases have received relatively limited attention in the financial literature, especially when compared to conference calls. This is surprising given that press releases are usually released first, are more widely disseminated, and are written in more structured and accessible language than conference call transcripts. Moreover, press releases are the documents that were most often targeted in high-profile cybersecurity breaches—such as the earnings press release hacking scandal documented by \citet{akey2022price}—underscoring their value in shaping market expectations.

This paper investigates whether and how the textual content of earnings press releases predicts earnings announcement date stock return responses. Using a comprehensive dataset of over 138,000 press releases issued between 2005 and 2023, we evaluate the predictive power of soft information using a variety of natural language processing (NLP) techniques. These include traditional bag-of-words models, specifically Latent Dirichlet Allocation (LDA), and contextual embedding models from the BERT family, including FinBERT—fine-tuned specifically for financial text. We construct real-time return forecasts using a rolling window approach that avoids look-ahead bias, and we assess the explanatory power of soft information embedded in press releases while controlling for earnings surprises using both regression and Shapley values.

Soft information extracted from press releases is just as informative as earnings surprises in explaining stock price reactions on announcement days. Among the models evaluated, FinBERT provides the most effective extraction of soft information from press releases for explaining stock returns, while LDA-based representations enhance interpretability by highlighting key thematic content. Since earnings announcements typically occur outside regular trading hours, we show that stock prices fully incorporate this soft information by market open, making it impossible for investors to profit from trading strategies based on publicly released press release signals—providing strong evidence for market efficiency. However, if this information becomes available before the scheduled release—as was the case in documented hacking incidents—traders can correctly identify high-return stocks with significantly better-than-random accuracy.

Our experiments evaluate five feature extraction methods, combining traditional bag-of-words and BERT-based models, to predict stock returns from earnings press releases. Our contributions are threefold: 
(1) Unlike conventional approaches based on static train-test splits, our framework supports online learning and adapts easily to new data;  
(2) We integrate the advantages of bag-of-words, BERT and large language models to enhance explainability by identifying key tokens and topics driving returns; 
(3) We show that earnings surprise and text are equally important in predicting announcement-day returns using linear models, achieving an $R^2\sim 4\%$ over 134k observations.
We further demonstrate through a trading strategy that: 
(1) the market is efficient, and investors cannot earn excess returns if they enter at market open following the release of overnight earnings news; 
(2) however, if earnings surprises and press release content are available in advance, investors can on-average correctly predict 56\% of the top 10 performing stocks using the agreement between hard and soft information.

\section{Related Work}
The Efficient Market Hypothesis (EMH) was formally tested and supported in the seminal paper by \citet{fama1969adjustment}, which showed that stock prices quickly incorporate publicly available information surrounding stock splits. However, \citet{ball1968} challenged this view by documenting a post-earnings announcement drift (PEAD), showing that stock prices continue to adjust after earnings are released — suggesting delayed incorporation of earnings news. This anomaly was further explored and substantiated by \citet{bernard1989post}, who provided a more systematic analysis of PEAD and emphasized the predictable nature of post-announcement returns. More recently, \citet{martineau2021rest} finds that markets have become increasingly efficient over time in processing earnings surprises (i.e., hard information), particularly due to improvements in information dissemination and algorithmic trading.  

However, much less is known about how markets process soft information—the qualitative content embedded in earnings press releases (i.e., soft information), such as managerial tone. Meaningful signals about future firm performance are found in press releases that do impact stock returns at the time of earnings announcements. For example, \citet{akey2022price} find that hackers who stole press releases ahead of earnings announcements relied on both earnings surprises and on the qualitative content of the press releases to inform their trading decisions.  \citet{akey2022price} use elastic net to capture the soft information embedded in the press releases. In this paper, we extend this work by using a mixture of models to capture the soft information in earnings press releases. 

A number of studies examine the content of earnings conference calls and how it relates to stock returns \citep[e.g.,][]{price2012earnings, meursault2023pead, garcia2023colour, ecc-analyzer}. However, only a few studies examined the content of earnings press releases \citep[e.g.,][]{huang2014tone, akey2022price}. Press releases are important because they represent the main piece of news that is issued by corporations at the time of the announcement that initiates the first price adjustment to earnings news. 
On the methodology front,  \citet{ecc-analyzer} employ multi-model analysis of conference call transcripts and audio to predict stock volatility, but their sample is limited to 500 firms in 2017. In contrast, we analyze press releases for over 6,000 unique stocks from 2006 to 2023, with time-varying return predictions and better interpretability. We further differentiate from \citet{ecc-analyzer}  since we examine the combination of both hard and soft information in how it impacts first moments, i.e,. stock returns, and how fast the information is impounded into stock prices.

\section{Data}

Our sample period begins in January 2005 and ends in December 2023. We obtain earnings dates and times, analysts' forecasts, and announced EPS from LSEG IBES Academic. We retrieve daily stock returns and open and closing prices from CRSP Daily Stock Files and intraday quotes from NYSE Trade and Quote (TAQ). 
% We focus solely on quarterly earnings announcements that are released after-hours (between 4 p.m.\ to 9.30 a.m.\ the following day) as more than 98\% occurs during that time \citep{gregoire2022earnings}. 
Following \cite{martineau2021rest}, we only select earnings announcements for which there is analyst following so that we compute earnings surprises (hard information) as 

\begin{equation}
	Surprise_{c,\tau} =  \frac{EPS_{c,\tau}-E_{\tau-1}[EPS_{c,\tau}]}{P_{c,\tau-5}},
\end{equation}

\noindent where $EPS_{c,\tau}$ is firm $c$'s earnings per share of quarterly earnings announcement announced on date $\tau$, and $E_{\tau-1}[EPS_{c,\tau}]$ is the expected earnings per share measured as the consensus analyst forecasts. The consensus analyst forecast is the median of all latest analyst forecasts issued over the 90 days before the earnings announcement date. We scale the surprise by the firm’s stock price $P_{c,\tau-5}$ five trading days before the announcement. %  and winsorize earnings surprises at the 1st and 99th percentiles; \\ winsorization is done later

We retrieve earnings press releases, in HTML format, published as attachments to 8-K filings from SEC Edgar.  We retrieve 158,797 earnings press releases for 6,543 unique firms (PERMNOs in CRSP).  After merging with CRSP daily stock returns, we have a total of 140,425 unique stock-earnings announcement observations. After preprocessing steps of earnings press releases (removing articles that are too short or too long), we have 138,676 stock-earnings announcement observations. Since 2005 is our initial training sample, our main analysis dataset contains 134,354 press releases from 2006 to 2023. 

We apply the following text preprocessing steps:
\begin{enumerate}
  \item Extract text within the \texttt{<body>} tag.
  \item Remove tables by decomposing \texttt{<table>} tags.
  \item Remove content common across press releases, such as ``forward-looking statements'', contact information, and ``non-GAAP financial measures disclosure''.
  \item Remove page/slide numbers, headers like ``Exhibit 99.x'', phone numbers, and phrases such as ``For immediate release.'' and ``For more information''. 
  \item Discard articles that are too short (<100 characters) or excessively long (>1,000,000 characters) either before or after cleaning. Fewer than 100 articles exceed 100,000 characters.
\end{enumerate}
Despite these efforts, variability in press release formats and the unstructured nature of text data introduce residual noise. Further, for bag-of-words vectorization, we apply standard text preprocessing: unescape HTML characters; remove URLs, email addresses, numbers, and special symbols; normalize whitespace and newlines; lemmatize tokens; and remove stop words and repeated patterns (\textit{e.g.}, ``year year'', ``month month'').

\begin{figure*}[htb]
  \centering
  \begin{subfigure}[b]{0.3\linewidth}
  \centering
  \includegraphics[width=\linewidth]{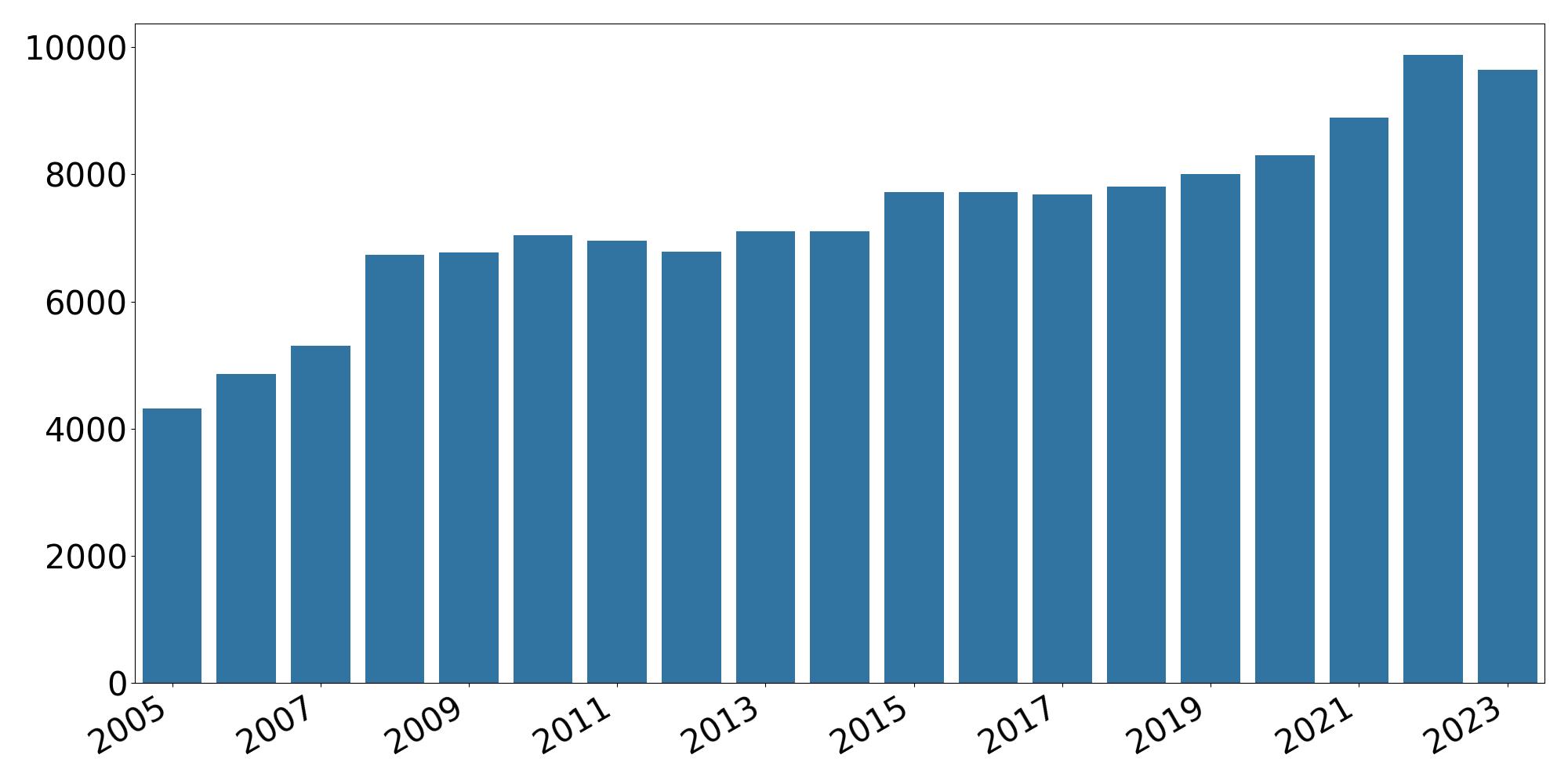}
  \caption{Total Number of Articles}
  \end{subfigure}
  \hfill
  \begin{subfigure}[b]{0.3\linewidth}
  \centering
  \includegraphics[width=\linewidth]{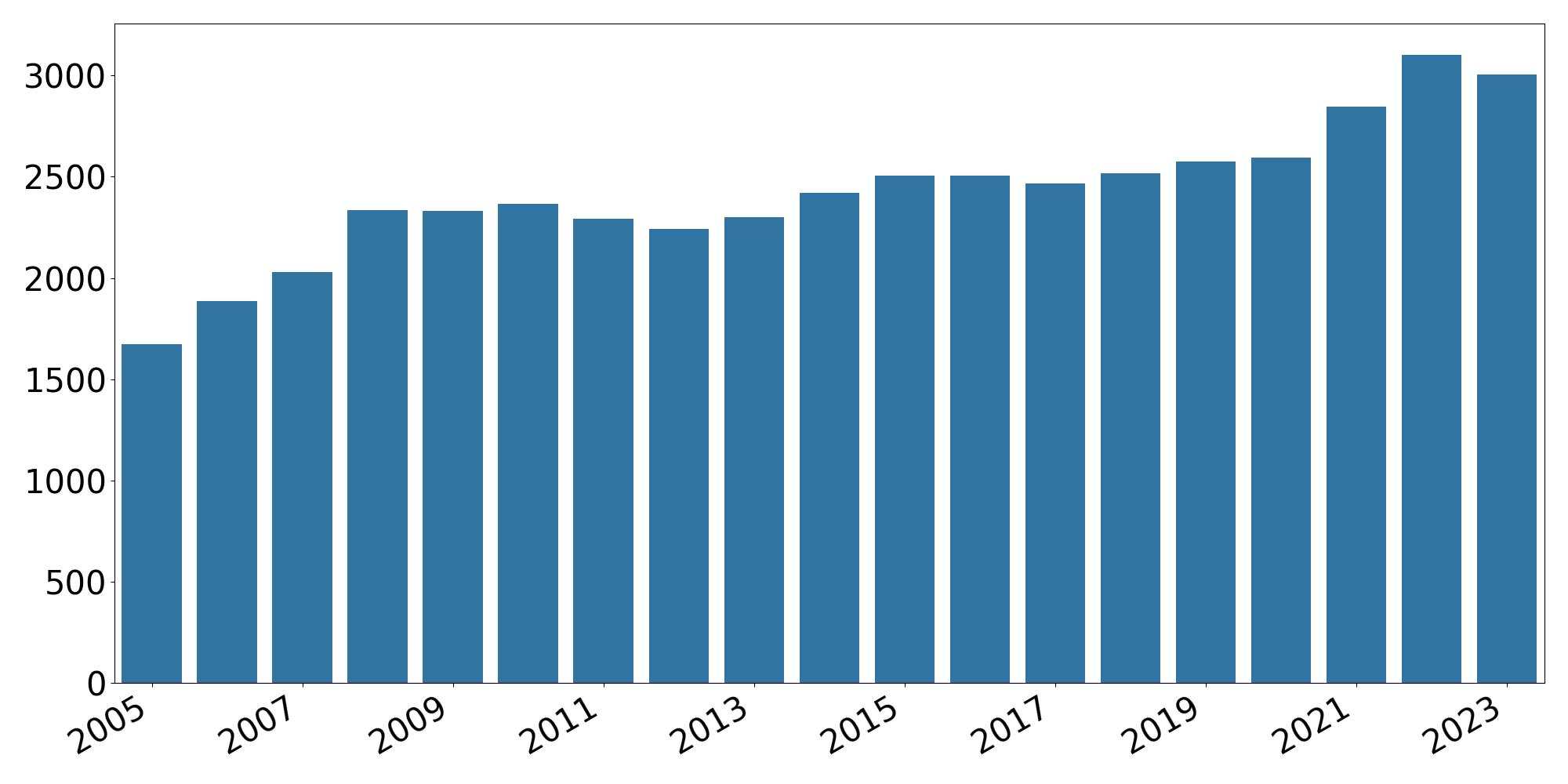}
  \caption{Total Number of Stocks}
  \end{subfigure}
  \hfill
  \begin{subfigure}[b]{0.3\linewidth}
  \centering
  \includegraphics[width=\linewidth]{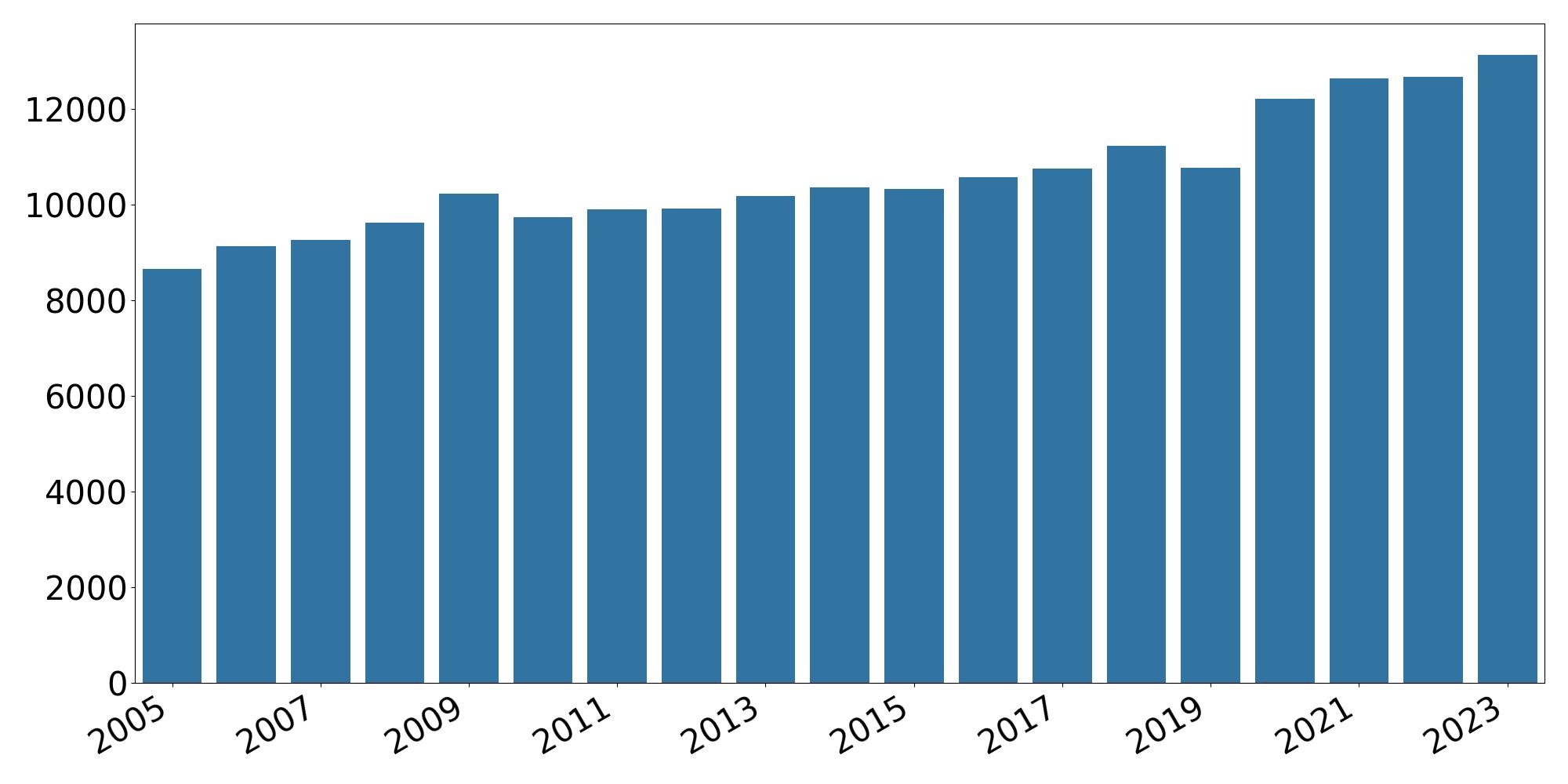}
  \caption{Average Number of Characters}
  \end{subfigure}
  \caption{General Statistics \label{fig:statistics}}
  \Description{General Statistics}
\end{figure*}

We retain press releases published after hours, \textit{i.e.}, on or after 4:00 p.m. or strictly before 9:30 a.m. ET. This comprises 97\% of the dataset. Announcement day returns are calculated as the percentage change in stock price from the prior close to the post-announcement close.
Figure~\ref{fig:statistics} summarizes the dataset: article count doubles from $\sim$5,000 to $\sim$10,000 between 2005 and 2023; approximately 2,500 U.S. stocks are covered across all market capitalizations per year\footnote{On average, there are 20$\sim$40 stocks traded per announcement day.}; and each article contains ~10,000 characters on average.

\section{Vectorization Techniques}
To construct numerical representations of the press release text, we use the following techniques from traditional bag-of-words approaches and BERT models, and compare their performance and interpretability.

\subsection{Bag-of-Words}
Instead of using basic count or TF-IDF representations, we apply Latent Dirichlet Allocation (LDA) to extract thematic content from the text, following \cite{bkmx}. LDA provides a lower-dimensional representation of the document-term matrix by modeling each article’s term counts as a multinomial distribution over topics, where each topic is a probability distribution over terms. Articles are then represented as mixtures of topics, capturing underlying themes.
For article $a$, and topic $i$, the topic attention vector $f_{i,a}$ is defined as
\begin{equation}\label{eq:topic-attention}
  f_{i,a} = \frac{\sum_{j=1}^{N_a}\mathbb{1}(\hat{z}_{j,a}=i)}{\sum_{q=1}^K\sum_{j=1}^{N_a}\mathbb{1}(\hat{z}_{j,a}=q)},
\end{equation}
where $\hat{z}_{j,a}$ is the topic assignment of word $j$, $N_a$ is the total vocabulary count in the article $a$, and $K$ is the total number of topics. This vector quantifies the degree to which article $a$ \textit{attends} to the topic $i$. 

We use the BKMX taxonomy from \cite{bkmx} as benchmark, which defines 180 topics across 3,592 phrases, spanning economics and politics. Each article is encoded as a 180-dimensional topic vector, and topics are grouped into hierarchical metatopics. In our experiments, we chose 11 metatopics of interest. However, the BKMX model, trained on \textit{Wall Street Journal} articles from 1984-2017, may not generalize well to press releases. 
It includes forward-looking tokens (\textit{e.g.} ``iphone'' in \textit{Mobile device}, ``twitter'' in \textit{Internet}), while missing newer terms (\textit{e.g.} ``president Biden''). Ambiguities also arise. Similar tokens may appear in multiple topics (\textit{e.g.} ``cit'' appears in both \textit{Buffett} and \textit{Major concerns}).

To address these limitations, we adopt online LDA (oLDA) \cite{olda, olda2}, training a dynamic topic model tailored to after-hours press releases. Cleaned articles are vectorized using a count vectorizer, then an oLDA model is incrementally trained year by year with parameters: learning offset $\tau_0=10$, decay $0.7$, document-topic prior $\alpha=1$, and topic-word prior $\eta=1$. We use the topic model trained on years $[2005, t-1]$ to vectorize the text in year $t$, minimizing the look-ahead bias in vocabulary and topic identification.
We use 50 topics with 10 key tokens each, representing each article as a 50-dimensional vector. To improve topic interpretability, we leverage GPT-4o-2024-08-06 \cite{gpt4o} for labelling topics and condensing topics into 10 metatopics. We find GPT-4o performs best with $\sim$500 input tokens, which ensures consistent outputs.
This setup captures terminology specific to earnings press releases and supports continual updates with new data (\textit{e.g.}, post-2024).
Still, like BKMX and BERT-based models, oLDA may assign the same token to multiple topics or repeatedly include similar tokens in one topic, limiting semantic distinctiveness.

\subsection{BERT Models}

Bi-directional Encoder Representations from Transformers (BERT) \cite{bert} is a transformer-based language model pre-trained on large text corpora and adaptable to downstream tasks through fine-tuning. We use several pre-trained models from the BERT family, BERT-base-uncased \cite{bert}, MPNET \cite{mpnet}, and FinBERT \cite{finbert}, to extract contextual embeddings for each article.\footnote{To minimize potential forward-looking bias in the training data, we focus on models developed prior to 2020.} Specifically, we represent each article as a 768-dimensional vector by applying mean pooling to the last hidden states of the first 512 tokens.
FinBERT is a domain-specific variant fine-tuned for financial sentiment analysis using the Reuters TRC2 corpus and Financial PhraseBank. 
While standard BERT models rely on masked language modeling, which captures token-level context but ignores inter-token dependencies, MPNET improves on this by incorporating permuted language modeling to better learn both positional and contextual relationships.
% As shown in Figure~\ref{fig:bert_similarity}, c
Cosine similarities between BERT and FinBERT embeddings are moderately high, with an average of $0.72\pm 0.03$, ranging from 0.6 to 0.8, indicating alignment in semantics. In contrast, embeddings from BERT and MPNET are mostly orthogonal, with a mean similarity of $0.01\pm 0.03$, suggesting they capture distinct features. This diversity in representations indicates that combining models may help capture complementary aspects of the text and improve downstream performance.

% \begin{figure}[!htb]
%   \centering
%   \includegraphics[width=\linewidth]{tables_25_05_01/bert_based_similarity.jpg}
%   \caption{BERT-based Embeddings Similarities \label{fig:bert_similarity}}
%   \Description{BERT-based Embeddings Similarities.}
% \end{figure}

\section{Analyzing Press Release Articles}

\subsection{Return Score Construction}

Let $X_{c,\tau}$ denote the vectorized press release for company $c$ on announcement day $\tau$, and let $\text{Ret}_{c,\tau}$ be the corresponding return. We use Lasso regression with weights $w$ to identify predictive features:
\begin{equation}\label{eq:lasso}
  \arg\min_w \frac{1}{2N} \|X_{c,\tau}w-\text{Ret}_{c,\tau}\|_2^2 + \lambda \|w\|_1,
\end{equation} 
with regularization parameter $\lambda=10^{-5}$. %Five-fold cross-validation is used to enhance robustness.
We adopt a rolling window framework: the model is trained on data from year $t$ and then used to generate return scores $\hat{\text{Ret}}_{c,\tau}=X_{c,\tau}w$ (denoted $\text{Soft}_{c,\tau}$) for year $t+1$. This process is repeated annually from 2005 to 2023, ensuring that the soft information capturing returns is updated with new text data while avoiding look-ahead bias.
To assess the predictive power of the return score, we run the following cross-sectional regression over the 2006–2023 period:
\begin{equation}\label{eq:regression}
	\text{Ret}_{c,\tau} = \alpha + \beta_0 \text{Surprise}_{c,\tau} + \beta_1 \text{Soft}_{c,\tau}+\epsilon,
\end{equation}
where $Ret_{c,\tau}$ corresponds to the announcement date stock return for firm $c$ on date $\tau$. Returns are computed from the 4 p.m. close preceding the announcement to 4 p.m.\ the next day. $Surprise_{c,\tau} $ corresponds to the earnings surprise that is winsorized at the 1st and 99th percentile. 
% $\text{Soft}_{c,\tau} $ is the soft information captured in earnings press releases. 
Standard errors are clustered by stock and announcement date. 

We further compute feature importance using SHAP values \cite{shapexact,SHAP}, with exact explainer and normalization:
\begin{align}
	I_j = \frac{1}{N} \sum_{i=1}^{N} \abs{\underset{S\subset X\setminus\bs{j}}{\EE}\closedb{y_{S\cup \bs{j}} - y_S}}, \quad \hat{I}_j = \frac{I_j}{\sum_{k=1}^n I_k} \times 100,
\end{align}
where $I_j$ is the mean absolute SHAP value for feature $j\in X$ across $N$ samples, representing the expected difference in the model outcomes $y$ when $j$ is added to a subset of features $S\subset X\setminus\bs{j}$, and $\hat{I}_j$ is normalized across all $n$ features.
Table~\ref{tab:regression1} summarizes the results. Column (1) reports that in an univariate analysis, $Surprise$ is positive and statistically significant at the 1\%. A one standard deviation (0.02) increase in the earnings surprise is associated with a 1.6\% (0.02 $\times$ 0.81 $\approx$ 1.6\%) increase in returns. The $R^2$ indicates that earnings surprise explains 3.3\% of the variation in announcement date returns. Columns (2)-(6) control for the soft information from press releases estimated from different models. Across all model specifications, Soft is positive and statistically significant at the 1\%, increasing the $R^2$ up to 4.4\%.\footnote{Surprise tends to bias the $R^2$ downward. It can be improved by using deciles of surprise instead of the actual level. The range of $R^2$ values we obtain is consistent with the literature on returns to surprise \cite{BKR19, martineau2021rest, akey2022price}.} The soft model estimated through FinBERT corresponds to the largest increase in $R^2$ relative to the univariate analysis in Column (1). Column (6) reports that a one standard deviation increase in Soft (0.016) is associated with an increase  of 1\% in returns (0.016 $\times$ 0.60 $\approx$ 1\%).

While earnings surprise dominates when using bag-of-words vectorization, the soft variable still contributes additional explanatory power. Notably, BERT-based models, especially FinBERT, can outperform earnings surprise in SHAP importance, highlighting their ability to extract meaningful return-predictive information from text. Columns (6) report a SHAP of 48\% for earnings surprises and 52\% for the Soft information.

\begin{table}[!htb]
\caption{Cross-Sectional Regression Analysis}\label{tab:regression1}

\centering

\resizebox*{\linewidth}{!}{\begin{tabular}{lcccccc}
\hline\hline
& &  \multicolumn{5}{c}{Soft independent variable}\\
 \cmidrule(lr){3-7}
 &  & BKMX & OLDA & BERT & MPNET & FINBERT \\
  & (1)  & (2) & (3) & (4) & (5) & (6)\\
\cmidrule(lr){2-2} \cmidrule(lr){3-7}
Surprise & 0.8056*** & 0.8001*** & 0.8030*** & 0.7714*** & 0.7891*** & 0.7579*** \\
 & (0.03) & (0.03) & (0.03) & (0.02) & (0.03) & (0.02) \\
Soft &  & 0.2402*** & 0.3242*** & 0.5602*** & 0.2309*** & 0.6006*** \\
 &  & (0.03) & (0.07) & (0.03) & (0.02) & (0.03) \\
Intercept & -0.0001 & 0.0000 & -0.0000 & 0.0002 & -0.0001 & 0.0003 \\
 & (0.00) & (0.00) & (0.00) & (0.00) & (0.00) & (0.00) \\
$R^2$ & 0.033 & 0.034 & 0.033 & 0.040 & 0.035 & 0.044 \\
$N$ & 134354 & 134354 & 134354 & 134354 & 134354 & 134354 \\
SHAP(Surprise) & 100.00 & 80.79 & 85.42 & 54.18 & 66.53 & 47.88 \\
SHAP(Soft) &  & 19.21 & 14.58 & 45.82 & 33.47 & 52.12 \\
\hline
\end{tabular}
}
\end{table}

Next, we evaluate the predictive power of soft information in predicting earnings announcement date returns when including all the constructed soft variables in the following regression:
\begin{equation}\label{eq:regression_combined}
  \text{Ret}_{c,\tau} = \alpha + \beta_0 \text{Surprise}_{c,\tau} + \sum_{I} \beta_I \text{Soft}_{c,\tau}^{I}+\epsilon.
\end{equation}
Table~\ref{tab:regression-combined} presents the results. As more soft variables are incorporated, the marginal contribution of earnings surprise declines. 

\begin{table}[!htb]
\caption{Combining Different Soft Variables for Announcement Day Return Prediction}\label{tab:regression-combined}

\centering

\resizebox*{\linewidth}{!}{\begin{tabular}{lccccc}
\hline\hline
 & (1) & (2) & (3) & (4) & (5) \\
\cmidrule(lr){2-5} \cmidrule(lr){6-6}
Surprise & 0.7583*** & 0.7580*** & 0.7580*** & 0.7578*** & 0.7573*** \\
 & (0.02) & (0.02) & (0.02) & (0.02) & (0.02) \\
Soft$^{BKMX}$ & -0.0372 &  &  &  & 0.0046 \\
 & (0.03) &  &  &  & (0.03) \\
Soft$^{OLDA}$ &  & -0.2774*** &  &  & -0.2970*** \\
 &  & (0.07) &  &  & (0.07) \\
Soft$^{BERT}$ &  &  & -0.0055 &  & 0.0180 \\
 &  &  & (0.03) &  & (0.04) \\
Soft$^{MPNET}$ &  &  &  & 0.0059 & 0.0183 \\
 &  &  &  & (0.02) & (0.02) \\
Soft$^{FINBERT}$ & 0.6071*** & 0.6288*** & 0.6043*** & 0.5972*** & 0.6075*** \\
 & (0.03) & (0.03) & (0.03) & (0.03) & (0.03) \\
Intercept & 0.0003 & 0.0003 & 0.0003 & 0.0003 & 0.0003 \\
 & (0.00) & (0.00) & (0.00) & (0.00) & (0.00) \\
$R^2$ & 0.044 & 0.044 & 0.044 & 0.044 & 0.044 \\
$N$ & 134354 & 134354 & 134354 & 134354 & 134354 \\
SHAP(Surprise) & 46.76 & 43.59 & 47.54 & 47.71 & 42.70 \\
SHAP(Soft$^{BKMX}$) & 1.82 &  &  &  & 0.21 \\
SHAP(Soft$^{OLDA}$) &  & 6.74 &  &  & 7.08 \\
SHAP(Soft$^{BERT}$) &  &  & 0.40 &  & 1.18 \\
SHAP(Soft$^{MPNET}$) &  &  &  & 0.64 & 1.78 \\
SHAP(Soft$^{FINBERT}$) & 51.42 & 49.67 & 52.06 & 51.65 & 47.05 \\
\hline
\end{tabular}
}
\end{table}

\begin{figure}[!htb]
	\centering
	\includegraphics[width=\linewidth]{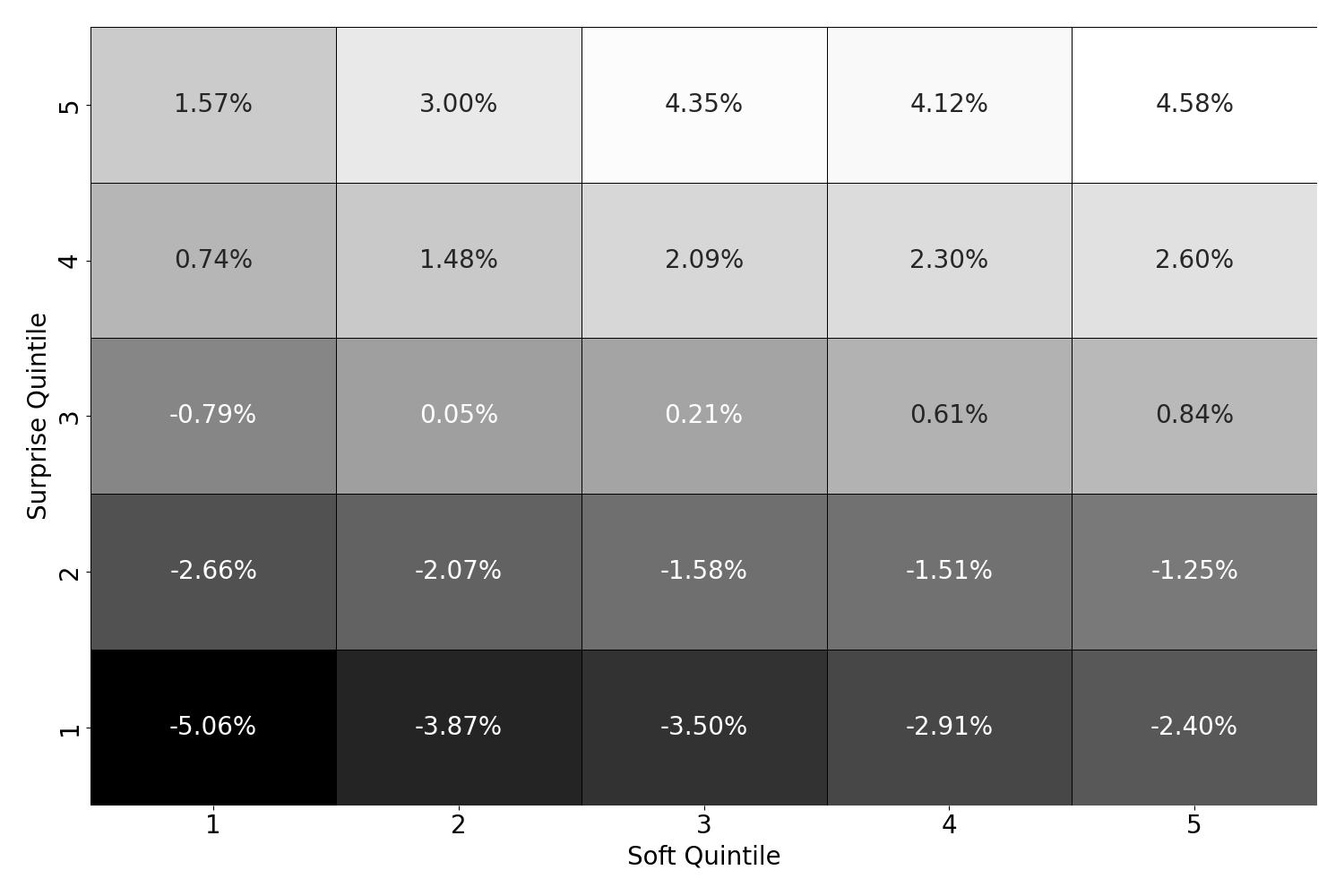}
	\caption{Average Returns Sorted by Surprise and Soft}\label{fig:avg-ret}
	\Description{Average return heatmap, sorted by soft and surprise}
\end{figure}

To further emphasize the importance of capturing soft information embedded in press releases to predict the price revelation on earnings announcement days, Figure~\ref{fig:avg-ret} shows the average return heatmap, sorted by quarterly quintiles of $\text{Soft}^{Mean}$ and realized earnings surprise. We define an aggregated textual signal, $\text{Soft}^{Mean}$, as the average of soft variables from BKMX, oLDA, BERT, MPNET, and FinBERT, to provide a unified measure of sentiment. The heatmap shows that the average return is driven by the composite signal, with the highest returns of 4.58\% in the upper-right corner and the lowest returns of -5.06\% in the bottom-left corner. The heatmap emphasizes that a trader that only processes the earning surprise (hard information) will fail to capture much of the variation in return that is driven by the soft information. 

Both Table~\ref{tab:regression1} and Table~\ref{tab:regression-combined} rely on the contemporaneously observed surprise, which may overstate its predictive relevance. To address this, we construct an out-of-sample surprise (OOS-Surprise) using the same rolling regression approach applied to soft variables: we regress $\text{Ret}_\tau$ on surprise in year $t$, and use the fitted model to predict return scores in year $t+1$. 
Table~\ref{tab:information-reaction} reports the results. The averaged soft variable $\text{Soft}^{Mean}$ remains statistically significant and delivers SHAP values on par with FinBERT and realized surprise, validating its effectiveness as a composite signal. As expected, the predictive strength of OOS-Surprise declines relative to its realized counterpart. 
Most importantly, in contrast to earnings surprises, Column (4) shows that our soft information measures fail to explain stock returns a day prior to the announcement. This implies that text-based information is not reflected in prices prior to release, suggesting that, on average, there is a low likelihood of press release leakage. Earnings surprises are positively related to pre-announcement returns, which is commonly associated with the fact that earnings surprises contain stale information since analysts do not revise their forecasts daily \citep{doyle2006, martineau2021rest}.

\begin{table}[!htb]
	\caption{OOS-Surprise and Prior Day Return Prediction}\label{tab:information-reaction}
	
	\centering
	
	\resizebox*{\linewidth}{!}{\begin{tabular}{lcccc}
\hline\hline
 &\multicolumn{3}{c}{$\text{Ret}_{\tau}$} & $\text{Ret}_{\tau-1}$\\ 
 & (1) & (2) & (3) & (4) \\
\cmidrule(lr){2-4} \cmidrule(lr){5-5} 
Surprise & 0.7726*** &  &  & 0.0245** \\
 & (0.02) &  &  & (0.01) \\
OOS-Surprise &  & 0.5038*** & 0.5088*** &  \\
 &  & (0.02) & (0.02) &  \\
Soft$^{FINBERT}$ &  & 0.6114*** &  & 0.0101 \\
 &  & (0.03) &  & (0.02) \\
Soft$^{Mean}$ & 0.8276*** &  & 0.8155*** &  \\
 & (0.05) &  & (0.05) &  \\
Intercept & 0.0001 & 0.0006 & 0.0004 & 0.0011** \\
 & (0.00) & (0.00) & (0.00) & (0.00) \\
$R^2$ & 0.041 & 0.036 & 0.032 & 0.000 \\
$N$ & 129811 & 129811 & 129811 & 129811 \\
SHAP(Surprise) & 51.64 &  &  & 62.58 \\
SHAP(OOS-Surprise) &  & 41.76 & 47.30 &  \\
SHAP(Soft$^{FINBERT}$) &  & 58.24 &  & 37.42 \\
SHAP(Soft$^{Mean}$) & 48.36 &  & 52.70 &  \\
\hline
\end{tabular}
}
\end{table}

\subsection{Topic Analysis}
To understand how textual content relates to stock returns, we perform topic analysis based on the features selected by Lasso regression. We categorize metatopics using three approaches. 

For the BKMX method, we rely on the original taxonomy, from which we chose 11 metatopics. In the oLDA approach, we use the 50 topics identified by the model and group them into 9 metatopics using GPT-4o. For BERT-based models, we start with extracting influential tokens in a manner analogous to bag-of-words techniques. Since Lasso regression identifies which of the 768 BERT features are positively or negatively associated with returns, we map these weights back to tokens. 
Let $E \in\RR^{512\times 768}$ be the BERT embedding matrix for the first 512 tokens in an article, and $w\in\RR^{768}$ be the Lasso weight vector. The importance score vector $IS\in \RR^{512}$ for each token is computed as $IS = E \cdot w$.
% \begin{equation}\label{eq:bert-importance}
%   IS = E \cdot w.
% \end{equation}
We then filter out uninformative tokens, including: (1) wordpiece tokens starting with \#\footnote{We could instead merge wordpiece tokens into their parent tokens, but the current extracted tokens are already informative.}; (2) punctuation, numeric and single-character tokens; and (3) BERT-specific tokens [CLS], [SEP], [PAD], [MASK], [UNK].
From the remaining tokens, we identify the top 5 positive and negative tokens per article, then group them into topics. This results in approximately 13,000 unique tokens from 2006 to 2023. These tokens are divided into chunks of 500 and labeled using GPT-4o, mapped to the oLDA metatopics due to its greater relevance to earnings-related content. 
Tokens that GPT-4o fails to categorize are assigned to a generic ``Other'' category and excluded from further analysis. In total, GPT-4o successfully categorized 3,219 tokens for BERT, 3,065 for MPNET, and 2,382 for FinBERT, representing around 20\% of all extracted tokens. Token counts per topic over all articles are then normalized to sum to 100\% within each model.
% We categorize metatopics using three approaches. 
% First, for the BKMX method, we rely on the original taxonomy, from which chose 11 metatopics. 
% Second, in the oLDA approach, we use the 50 topics identified by the model and group them into 9 metatopics using GPT-4o. 
% Finally, for BERT-family models (BERT, MPNET, and FinBERT), we extract the top positive and negative tokens per article, resulting in approximately 13,000 unique tokens from 2006 to 2023. These tokens are divided into chunks of 500 and labeled using GPT-4o, mapped to the oLDA metatopics due to its greater relevance to earnings-related content. 
% 

\begin{figure*}[!htb]
  \centering
  \begin{subfigure}[b]{0.48\linewidth}
  \centering
  \includegraphics[width=\linewidth]{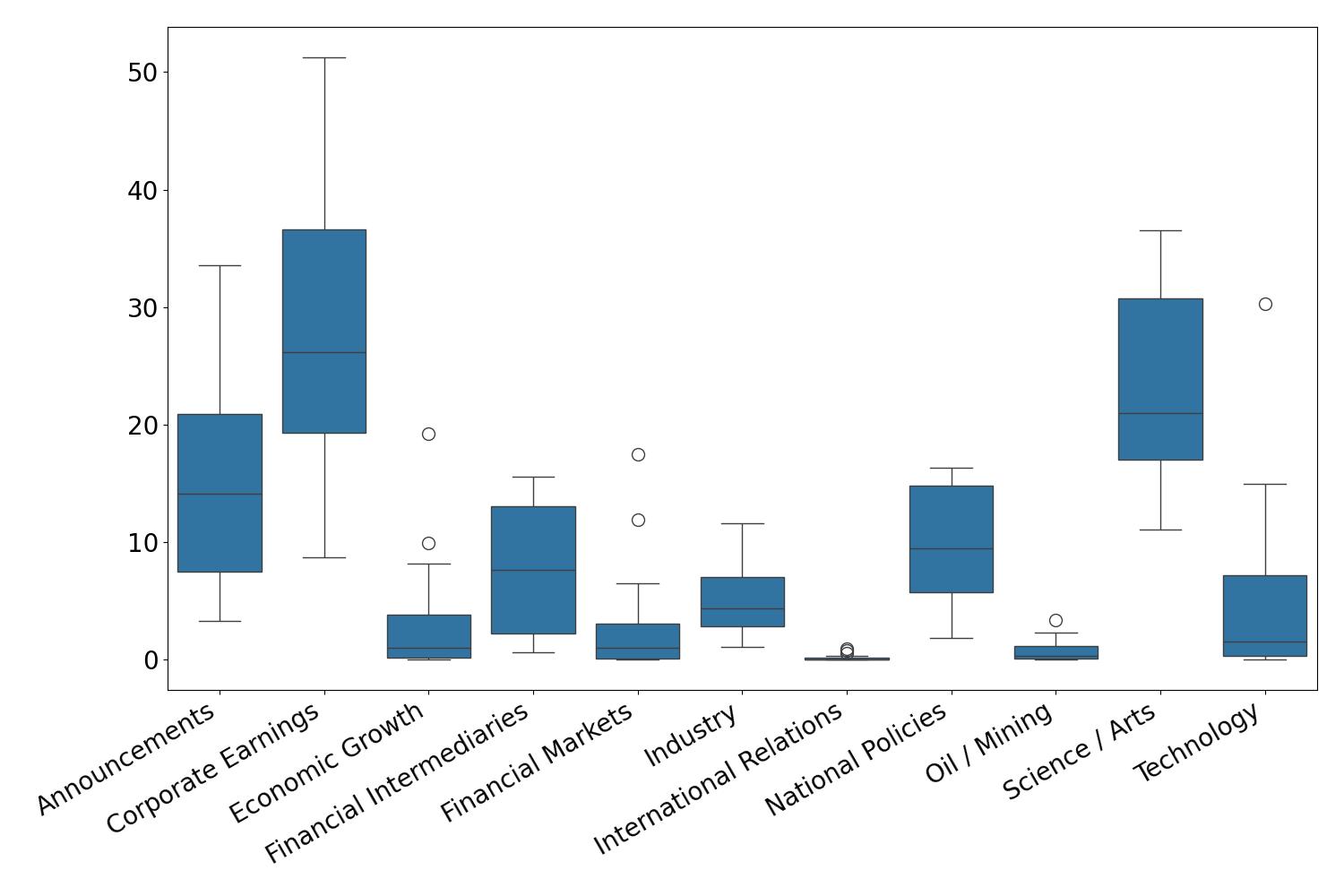}
  \caption{BKMX}
  \end{subfigure}
  \hfill
  \begin{subfigure}[b]{0.48\linewidth}
  \centering
  \includegraphics[width=\linewidth]{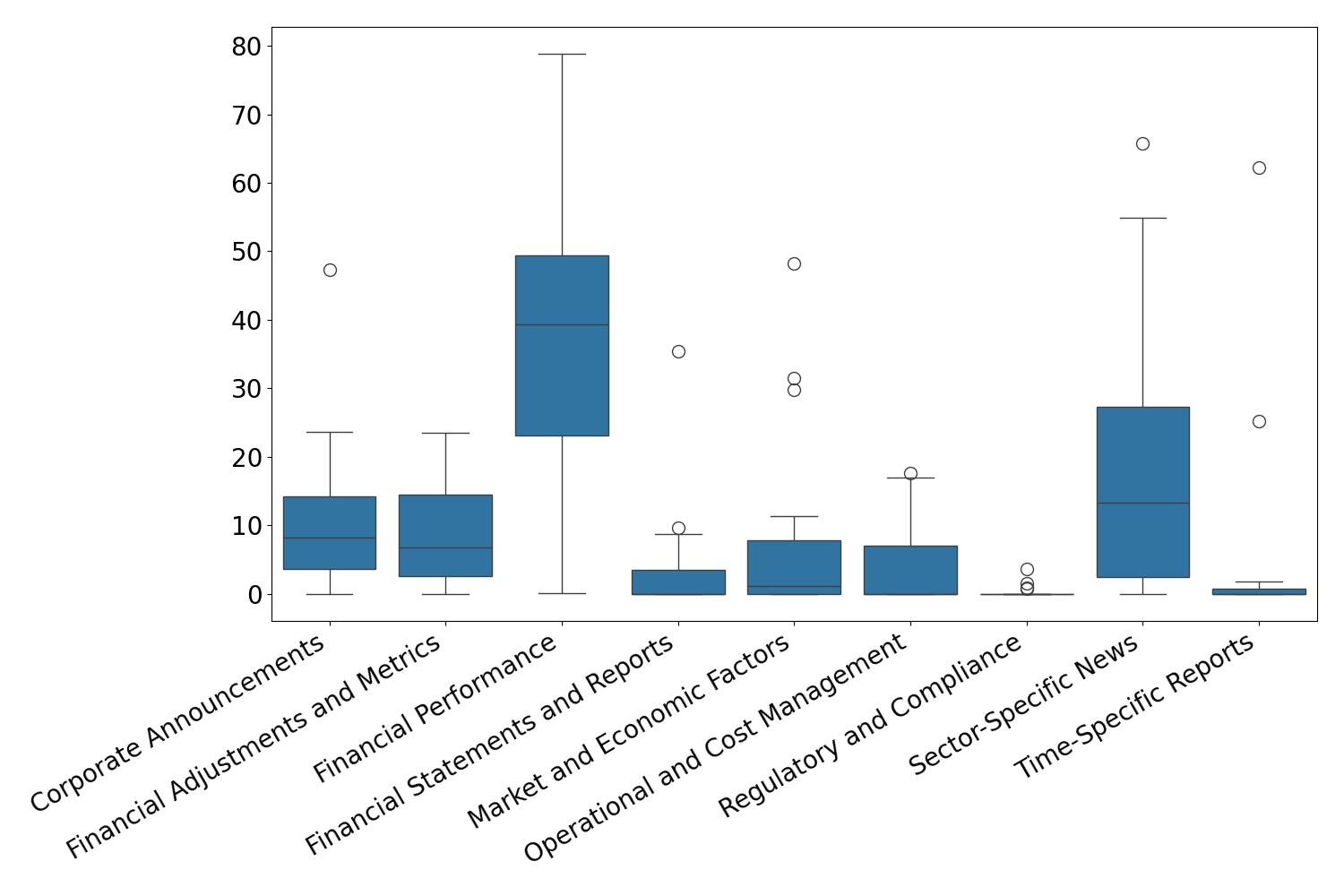}
  \caption{oLDA}
  \end{subfigure}
  \vspace{-0.3cm}
  \caption{Metatopic Explained Variance}\label{fig:metatopic-explained-variance}
  \Description{Metatopic Explained Variance for BKMX and oLDA}
\end{figure*}

\begin{figure*}[!htb]
  \centering
  \begin{subfigure}[b]{0.48\linewidth}
  \centering
  \includegraphics[width=\linewidth]{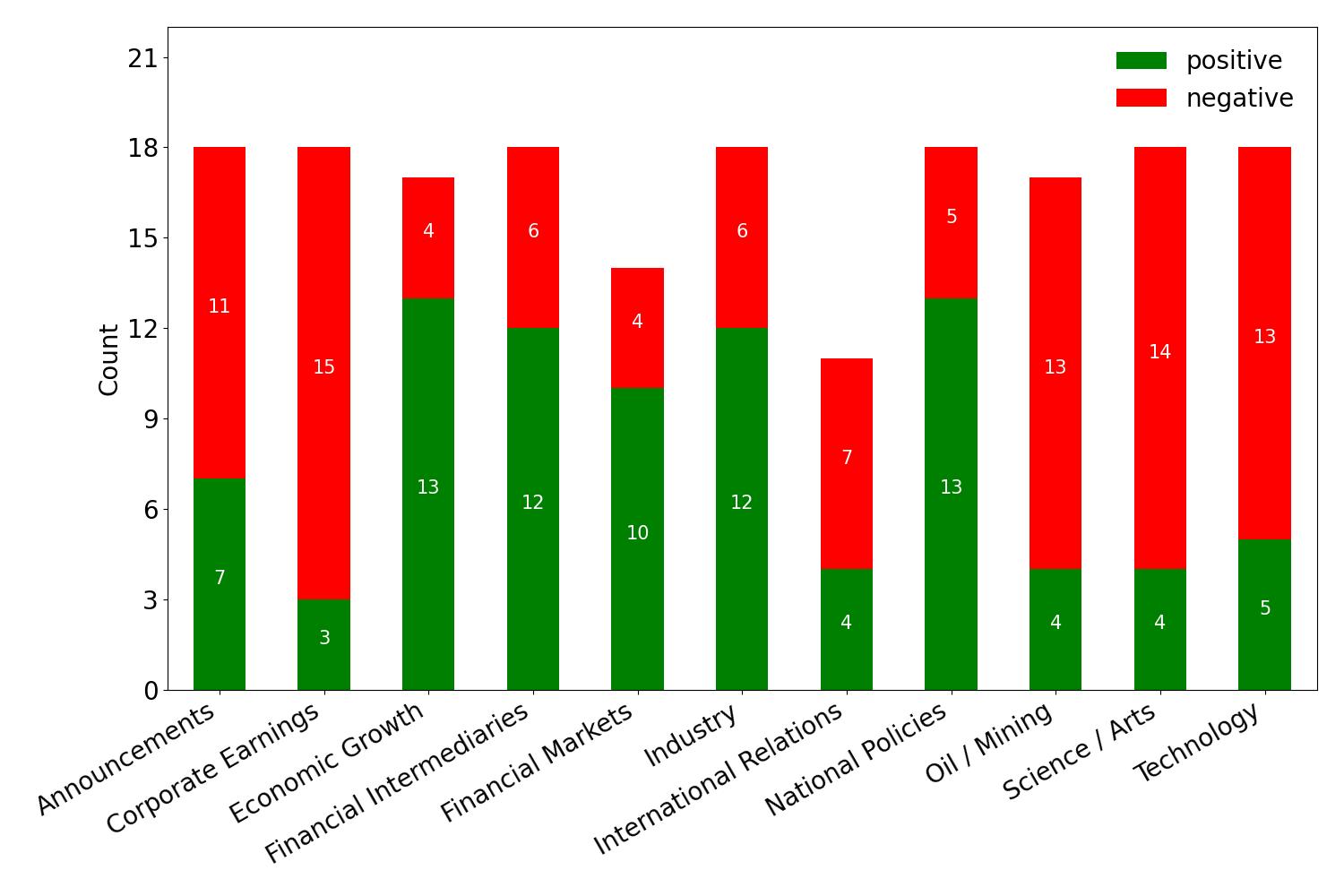}
  \caption{BKMX}
  \end{subfigure}
  \hfill
  \begin{subfigure}[b]{0.48\linewidth}
  \centering
  \includegraphics[width=\linewidth]{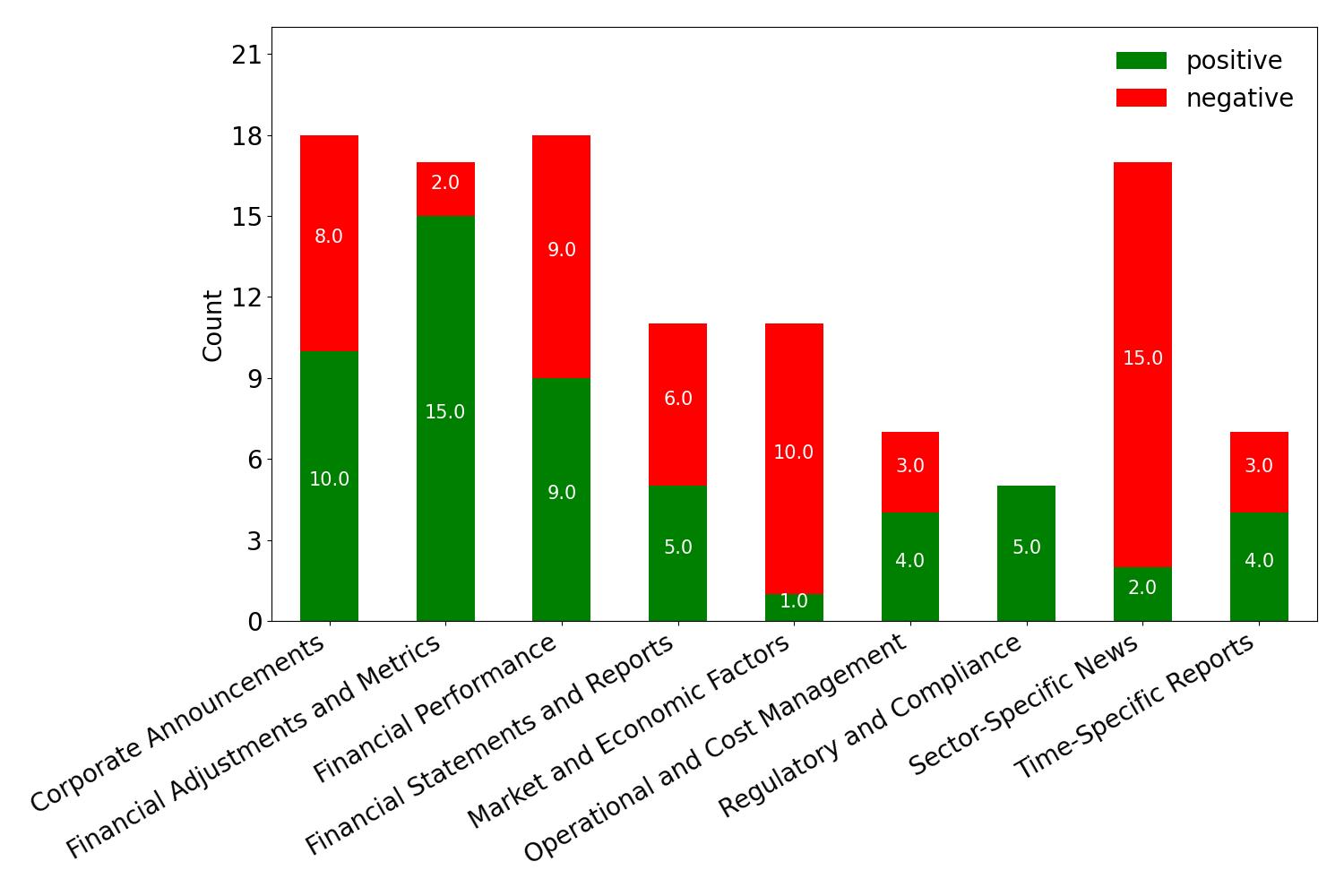}
  \caption{oLDA}
  \end{subfigure}
  \vspace{-0.3cm}
  \caption{Metatopic Weight Polarity}\label{fig:metatopic-weight-polarity}
  \Description{Metatopic Weight Polarity for BKMX and oLDA}
\end{figure*}

For both BKMX and oLDA, we compute the total weight of each metatopic $w_{M}$ by summing up the weights $w_k$ of all topics within the metatopic. We also calculate the explained variance $h(M)$ of each metatopic for each year from 2006 to 2023, using the formula:
\begin{align*}
  w_{M} &= \sum_{k\in M} w_k\\
  h(M) &= \frac{\sum_{i\in M}\sum_{j\in M} w_i w_j \text{Cov}(f_i, f_j)}{\sum_M\sum_{i\in M}\sum_{j\in M} w_i w_j \text{Cov}(f_i, f_j)} \times 100,
\end{align*}
where $f_i$, $f_j$ are topic attention vectors across all articles in year $t$ (Eq.~\eqref{eq:topic-attention}), and $w_i, w_j$ are fitted weights from year $t-1$. 
Figure~\ref{fig:metatopic-explained-variance} presents the distribution of explained variances, while Figure~\ref{fig:metatopic-weight-polarity} shows the polarity (positive vs. negative) of topic weights from 2005 to 2022.

Among the BKMX metatopics, the most influential are \textit{Announcements}, \textit{Corporate Earnings}, \textit{Financial Intermediaries}, \textit{National Politics}, and \textit{Science / Arts}. 
Notably, the \textit{Science / Arts} category includes topics such as ``challenges'', ``marketing'', and ``key role'' which, despite being misclassified in the original BKMX taxonomy, are closely related to corporate earnings and carry high regression weights. The \textit{Economic Growth} metatopic, although less influential in variance, shows stronger relevance during recessionary periods and years of elections, specifically 2007, 2008, 2011, and 2015, where its negative weights align with downturns in market performance and political cycle. In 2020, the weight is zero, which means that the weights of associated topics (recession \& growth) cancel out.

For oLDA, the most informative metatopics include \textit{Financial Performance}, covering topics: ``Year-End Results'', ``Quarterly Income'', ``Net Income'', ``Monthly Results'', ``Net Loss'', ``Revenue Growth'', ``Sales Growth'', ``Operating Margin'', ``Quarterly Report'', and ``Prior Year Comparison'';
as well as \textit{Sector-Specific News}, which includes domain-focused topics: ``Cancer Research'', ``Retail Sales'', ``Energy Utilities'', ``Oil and Gas'', ``Banking'', ``Insurance'', ``Technology Solutions'', ``Healthcare Services'', ``Digital Media'', and ``Hospitality''. Figure \ref{fig:metatopic-weight-polarity} highlights an interesting narrative employed by companies when discussing earnings results in press releases. 
When companies surpass earnings expectations or achieve positive results, Figure \ref{fig:metatopic-weight-polarity} shows that managers speak much more positively about financial adjustments, metrics, and performance than sector-specific news. When achieving below-than-expectations, managers speak much more negatively about sectors. This finding relates to the ``self-serving bias,'' where individuals attribute their successes to internal factors (e.g., managerial abilities) and failures to external factors (e.g., bad luck, others' actions) \citep{zuckerman1979attribution} and to managerial overconfidence \citep{doukas2007acquisitions}.

% We analyze token-level topic assignments for BERT-family models using oLDA taxonomy over 2006–2023. 
Table~\ref{tab:bert-token-classification} reports the classification results for BERT-family models.
% showing that BERT models, particularly FinBERT, align with oLDA topic assignments that are specific to earnings-related content.
As a fine-tuned model on financial text, FinBERT captures more tokens in \textit{Financial adjustments and metrics} (11.09\%) and \textit{Financial performance} (16.07\%), both associated with positive return signals ($IS > 0$). These topics focus on internal factors.
Conversely, topics such as \textit{Market and Economic Factors} (7.94\%), \textit{Time-Specific Reports} (7.11\%), \textit{Financial Performance} (6.8\%) and \textit{Sector-Specific News} (6.27\%) are linked to negative return signals ($IS < 0$), reflecting external factors like macroeconomic conditions, sector trends, and seasonality.
These results support the presence of a self-serving bias: internal factors are framed more positively than external ones. 
FinBERT's topic alignment with oLDA also suggests explainability of FinBERT in analyzing financial texts. However, a substantial fraction of tokens remains misclassified by GPT-4o, and BERT and MPNET, the non-financial specific models, are more difficult to interpret. Future works can improve token categorization through refined prompting strategies or pre-filtering methods.

% \begin{table*}[!htb]
% \caption{BERT Classification: BKMX Metatopic Classification}\label{tab:bert-classification-BKMX}

% \centering

% \resizebox*{\linewidth}{!}{\input{tables_25_05_01/bert_classification_BKMX.tex}}
% \end{table*}
\hspace{-0.25cm}

\begin{table*}[!htb]
\caption{BERT Token Classification}\label{tab:bert-token-classification}
% \vspace{-0.3cm}
\centering

\resizebox*{0.95\linewidth}{!}{\begin{tabular}{llccccccccc}
\toprule
Model & Sign & Corp. Ann. & Fin. Metrics & Fin. Perf. & Fin. Reports & Market / Econ. Factors & Op. / Cost Mgmt & Regulatory & Sector News & Time-Spec. Reports \\
\cmidrule(lr){1-2} \cmidrule(lr){3-11}
BERT & Pos & 0.58\% & 5.75\% & 7.96\% & 11.78\% & 6.36\% & 11.75\% & 1.29\% & 7.95\% & 6.76\% \\
 & Neg & 0.24\% & 2.34\% & 6.54\% & 3.28\% & 5.47\% & 10.11\% & 0.79\% & 5.83\% & 5.22\% \\
 MPNET & Pos & 0.90\% & 4.63\% & 5.53\% & 4.41\% & 3.37\% & 6.12\% & 0.43\% & 4.82\% & 17.48\% \\
 & Neg & 0.59\% & 6.75\% & 7.87\% & 2.62\% & 5.18\% & 9.53\% & 1.01\% & 9.13\% & 9.63\% \\
 FinBERT & Pos & 1.46\% & 11.09\% & 16.07\% & 4.43\% & 5.22\% & 5.11\% & 1.49\% & 6.82\% & 7.73\% \\
 & Neg & 1.85\% & 2.83\% & 6.80\% & 2.07\% & 7.94\% & 4.50\% & 1.22\% & 6.27\% & 7.11\% \\
\bottomrule
\end{tabular}
}
\end{table*}

\section{Applications}

We next examine two key applications of our method in capturing soft information embedded in earnings press releases. First, we investigate the speed at which soft information is incorporated into stock prices. Second, we quantify the improvement in predicting top/bottom-performing stocks by combining earnings press release information with earnings surprises if they are obtained in advance.

\subsection{Market Efficiency}\label{sec:mkt-efficiency}
In an ideal market without friction, all available information is instantly reflected in asset prices, leaving no room for investors to earn abnormal returns based on public disclosures alone \citep{efficient-market, fama1969adjustment}. To evaluate whether this holds for earnings press releases, we conduct a trading-based test of market efficiency.

On each day $\tau$, for each firm $c$, we predict the returns using all five models, and compute $\text{Soft}^{Mean}$ the average across the five model predictions. We then implement a long-short strategy based on the agreement between $\text{Soft}^{Mean}$ and earnings surprise.
We enter trades at 9:45 a.m since bid-ask spreads are wide at the opening of markets \citep{mcinish1992analysis, gregoire2022earnings}, and exit at the 4:00 p.m close. Using TAQ data, we exclude stocks with excessive illiquidity—defined as a relative bid-ask spread exceeding 20\% of the midquote:
\begin{equation}
  \frac{\abs{\text{Ask}-\text{Bid}}}{\text{Midquote}}\geq 0.2, \text{ where } \text{Midquote}=\frac{\text{Ask}+\text{Bid}}{2}.
\end{equation}
This filter removes 0.82\% of the data properly aligned in TAQ and event data. Among the remaining data, 54.25\% show consistent signals between surprise and $\text{Soft}^{Mean}$.
We long stocks $L$ with both positive surprise and positive $\text{Soft}^{Mean}$, and short stocks $S$ with both negative signals. To balance the portfolio, we trade $\min\bs{|L|,|S|}$ stocks, ranked by either surprise or $\text{Soft}^{Mean}$.
The portfolio is weighted by market capitalization on day $\tau-1$\footnote{Market-cap weighting yields more stable results than equal weighting.}. We consider two strategies $j\in \bs{1,2}$:
\begin{enumerate}
  \item Strategy 1 ($j=1$): Enter at ask (bid) if the signal is positive (negative), and exit at the closing price. $\text{Ret}_{\tau}^{OC,1} = \frac{p_\tau}{p_\tau^{o,k}}-1$, where $p_{\tau}$ is the closing price on day $\tau$, $p_\tau^{o,k}$ is the best ask/bid price at 9:45 a.m on day $\tau$, and $k\in \bs{\text{Ask}, \text{Bid}}$.
  \item Strategy 2 ($j=2$): Enter and exit using ask/bid quotes, capturing transaction costs. For instance, a long position buys at ask and sells at bid.  $\text{Ret}_{\tau}^{OC,2}$ is computed in a similar fashion, with $p_\tau$ the closing price replaced by ask/bid price at close. 
\end{enumerate}
Let $\text{Ret}_t^{OC,L,j}$ and $\text{Ret}_t^{OC,S,j}$ denote the average long and short returns on day $t$, weighted by market capitalization. The long-short return is defined as
\begin{equation}
  \text{LS}_t^j = \text{Ret}_t^{OC,L,j} - \text{Ret}_t^{OC,S,j}
\end{equation}

We test for abnormal returns by regressing $\text{LS}_t^j$ on constant and market factors:
\begin{align}
  \text{LS}_t^j &= \alpha, \label{eq:ls-alpha}\\
  \text{LS}_t^j &= \alpha + \beta_0 \text{Mkt-rf}_t + \beta_1 \text{HML}_t + \beta_2 \text{SMB}_t + \varepsilon_t \label{eq:ls-market},
\end{align}
where $\text{Mkt-rf}_t$ is the market return on day $t$ less risk-free rate, and $\text{HML}_t$ and $\text{SMB}_t$ are HML and SMB factors of \cite{fama1992cross}, respectively. 
The estimated $\alpha$ values from Eq.\eqref{eq:ls-alpha} are as follows: for strategy 1, $\alpha = -0.0004$ when returns are sorted by Surprise and $\alpha = -0.0007$ (statistically significant at the 5\% level) when sorted by Soft. For strategy 2, the corresponding values are $\alpha = -0.0055$ and $\alpha = -0.0051$, both significant at the 1\% level. 
The inclusion of the Mkt-RF, HML, and SMB factors has little impact on the estimated $\alpha$s. Although the returns are statistically significant, they are negative and close to zero, indicating that the trading strategy primarily incurs transaction costs and does not generate sustained profits. These findings are consistent with market efficiency.

\subsection{``Hacking'' the Earnings News}

In \cite{akey2022price} the authors document a case where hackers infiltrated US-based newswire services--notably PR Newswire, Business Wire, and Marketwired--and stole thousands of yet-to-be-published earnings press releases. The hackers sold this information to traders, who executed front-running trades just before the official release of the earnings announcements. This hacking operation occurred over several years (roughly 2010 to 2015) and the traders pocketed more than 100 million US dollars. Eventually, the SEC and DOJ charged more than 30 individuals in one of the largest securities fraud cases of its kind. The incident highlights that press releases are a critical node in the information dissemination chain, and any breach undermines investor confidence, market efficiency, and regulatory fairness.

\citet{akey2022price} find that it was crucial for traders to use both the content of the press releases and the earnings surprises to maximize their expected profits from these illegal activities. However, even when combining both signals, traders did not always achieve the highest expected returns. This is because earnings announcements are complex pieces of information that are not trivial to interpret, and they are often followed by conference calls between corporate managers and analysts that can substantially impact prices.  

Our method and results so far have demonstrated how important are earnings surprises and press releases in explaining announcement date returns, but if one ranks stocks based on expected returns and evaluate the ranking relative to the ground truth (\textit{i.e.}, realized returns), what fraction can we correctly identify? 

We select the days with at least 20 earnings announcements (1497 days), which constitutes for 111,104 events in total.
After filtering on soft/surprise agreement, there are 60,192 stocks remaining. Let $C$ be the number of stocks in top $k$ sorted by soft information or surprise that are also in top $k$ sorted by the realized announcement day return. We then compute the Precision@$k$ score as $\text{P@}k = \frac{\text{C}}{k}$. Table~\ref{tab:application2} shows the result based on surprise or soft information alone, while Table~\ref{tab:application2-filtered} shows the result with the agreement between the two signals. Comparing both tables reveals that integrating soft information from press releases with earnings surprises substantially enhances the ability to identify the top 10 best- and worst-performing stocks. Specifically, for P@10, the precision when using surprise or soft information separately to identify the top positive and negative return stocks ranges from 0.25 to 0.36. When combining both signals, P@10 increases significantly to a range of 0.44 to 0.52. These findings demonstrate the value of combining hard and soft information for predicting extreme stock performance and highlight that newswire companies must continue investing in cybersecurity to fight against data breaches \cite{bloomberg-hack}.

\begin{table}[!htb]
\caption{Precision with Separate Use of Surprise or Soft}\label{tab:application2}

\centering

\resizebox*{\linewidth}{!}{\begin{tabular}{llccccc}
\toprule
 &  & P@1 & P@2 & P@3 & P@5 & P@10 \\
\cmidrule(lr){3-7}
Top Positive & Surprise & 0.0908 & 0.1353 & 0.1710 & 0.2355 & 0.3472 \\
 & Soft$^{Mean}$ & 0.0371 & 0.0614 & 0.0853 & 0.1335 & 0.2453 \\
%  & Soft$^{BKMX}$ & 0.0274 & 0.0464 & 0.0688 & 0.1086 & 0.2185 \\
%  & Soft$^{OLDA}$ & 0.0267 & 0.0444 & 0.0623 & 0.1062 & 0.2153 \\
%  & Soft$^{BERT}$ & 0.0200 & 0.0418 & 0.0677 & 0.1137 & 0.2298 \\
%  & Soft$^{MPNET}$ & 0.0307 & 0.0544 & 0.0746 & 0.1193 & 0.2277 \\
%  & Soft$^{FINBERT}$ & 0.0274 & 0.0458 & 0.0677 & 0.1177 & 0.2331 \\
%  & OOS-Surprise & 0.0902 & 0.1356 & 0.1706 & 0.2357 & 0.3472 \\
Top Negative & Surprise & 0.0935 & 0.1403 & 0.1746 & 0.2379 & 0.3578 \\
 & Soft$^{Mean}$ & 0.0530 & 0.0855 & 0.1147 & 0.1633 & 0.2714 \\
%  & Soft$^{BKMX}$ & 0.0381 & 0.0605 & 0.0882 & 0.1300 & 0.2333 \\
%  & Soft$^{OLDA}$ & 0.0294 & 0.0478 & 0.0750 & 0.1190 & 0.2287 \\
%  & Soft$^{BERT}$ & 0.0521 & 0.0868 & 0.1124 & 0.1617 & 0.2695 \\
%  & Soft$^{MPNET}$ & 0.0407 & 0.0735 & 0.1044 & 0.1546 & 0.2566 \\
%  & Soft$^{FINBERT}$ & 0.0648 & 0.1052 & 0.1343 & 0.1768 & 0.2824 \\
%  & OOS-Surprise & 0.0929 & 0.1393 & 0.1741 & 0.2381 & 0.3578 \\
\bottomrule
\end{tabular}
}

\end{table}
\vspace{-0.5cm}
% \begin{table}[!htb]

\begin{table}[!htb]
\caption{Precision with Agreement in Surprise and Soft}\label{tab:application2-filtered}

\centering

\resizebox*{\linewidth}{!}{\begin{tabular}{llccccc}
\toprule
 &  & P@1 & P@2 & P@3 & P@5 & P@10 \\
\cmidrule(lr){3-7}
Top Positive & Surprise & 0.1242 & 0.2017 & 0.2594 & 0.3455 & 0.5170 \\
 & Soft$^{Mean}$ & 0.0698 & 0.1199 & 0.1689 & 0.2574 & 0.4411 \\
%  & Soft$^{BKMX}$ & 0.0615 & 0.0992 & 0.1467 & 0.2303 & 0.4098 \\
%  & Soft$^{OLDA}$ & 0.0514 & 0.0962 & 0.1383 & 0.2208 & 0.4035 \\
%  & Soft$^{BERT}$ & 0.0541 & 0.1022 & 0.1512 & 0.2472 & 0.4393 \\
%  & Soft$^{MPNET}$ & 0.0701 & 0.1182 & 0.1630 & 0.2528 & 0.4368 \\
%  & Soft$^{FINBERT}$ & 0.0595 & 0.1019 & 0.1543 & 0.2478 & 0.4401 \\
%  & OOS-Surprise & 0.1222 & 0.2017 & 0.2596 & 0.3456 & 0.5170 \\
Top Negative & Surprise & 0.1409 & 0.2111 & 0.2710 & 0.3606 & 0.5192 \\
 & Soft$^{Mean}$ & 0.0983 & 0.1638 & 0.2155 & 0.2979 & 0.4661 \\
%  & Soft$^{BKMX}$ & 0.0681 & 0.1283 & 0.1708 & 0.2514 & 0.4215 \\
%  & Soft$^{OLDA}$ & 0.0548 & 0.1089 & 0.1521 & 0.2389 & 0.4163 \\
%  & Soft$^{BERT}$ & 0.1122 & 0.1727 & 0.2325 & 0.3150 & 0.4810 \\
%  & Soft$^{MPNET}$ & 0.1009 & 0.1680 & 0.2224 & 0.2946 & 0.4678 \\
%  & Soft$^{FINBERT}$ & 0.1169 & 0.1934 & 0.2445 & 0.3269 & 0.4907 \\
%  & OOS-Surprise & 0.1369 & 0.2118 & 0.2705 & 0.3606 & 0.5192 \\
\bottomrule
\end{tabular}
}
\end{table}

\section{Conclusion}

In this paper, we analyze earnings press releases with a mixture of models to predict announcement day returns. We find that both earnings surprise (hard information) and press release text (soft information) explain a similar share of return variation. While FinBERT achieves the highest SHAP values and $R^2$, traditional bag-of-words models, especially oLDA, improve its predictions and offer greater explainability.
We also examine a long-short strategy based on hard and soft information. Although the strategy yields no profit when executed after market open, supporting market efficiency, it becomes profitable if information leaks the day before, underscoring the value of earnings announcements. 
Our future work will integrate conference calls texts and audios to explore information flow in greater depth.

% \section{Supporting Code}
% Tables and Figures can be replicated using the \href{https://mega.nz/folder/zkkmzLIS#-HPj2mm3r-qYA3ypu_6XfA}{supporting code}. Full code base will be available on GitHub after acceptance.

\begin{acks}
This research was carried out at Rotman School of Management, FinHub Financial Innovation Lab, University of Toronto. We gratefully acknowledge financial support from the Social Sciences and Humanities Research Council (SSHRC) of Canada grant number 435-2022-0745.
\end{acks}

\bibliographystyle{ACM-Reference-Format}
\bibliography{main}

% \appendix
% \section{Code Availability}
% We provide the full replication package and a experimental colab notebook for simple analysis on Google Drive: . All relevant GPT prompts and generated topic classifications are included. Full code base will be available on GitHub after acceptance.

\end{document}